\documentclass[12pt]{article}  

\usepackage{amsmath,amssymb,amsfonts}
\usepackage{longtable}

\def\mylist{\begin{list}{}{\setlength{\leftmargin}{0.5in}
               \setlength{\listparindent}{-0.5in}
               \setlength{\itemindent}{\listparindent}}}

\newcommand{\ket}[1]{|#1\,\rangle}

\newcommand{\nD}[1]{\not D}
\begin{document}

\begin{titlepage}

\title{A note on ${\cal N}\ge 6$ superconformal quantum field theories in three dimensions}
\author{Denis Bashkirov\\ {\it California Institute of Technology}}

\maketitle

\abstract{Based on the structure of the three-dimensional superconformal algebra we show that every irreducible ${\mathcal N}=6$ three-dimensional superconformal theory containes exactly one conserved $U(1)$-symmetry current in the stress tensor supermultiplet and that superconformal symmetry of every ${\mathcal N}=7$ superconformal theory is in fact enhanced to ${\mathcal N}=8$. Moreover, an irreducible ${\cal N}=8$ superconformal theory does not have any global symmetries. The first observation explains why all known examples of ${\mathcal N}=6$ superconformal theories have a global abelian symmetry.}

\end{titlepage}

\section{Introduction}

Conformal quantum field theories are not only an interesting special case of general quantum field theories. They are essential part of the definition of any quantum field theory as a relevant perturbation of an ultra-violet fixed point, that is, a conformal field theory. Superconformal field theories are needed to define supersymmetric quantum field theories. Beside this, the infrared limit of a quantum field theory is controlled by another conformal field theory which is called the infrared fixed point.

The simplest cases of (super)conformal field theories are free field theories without dimensionful parameters. These are always explicitly described in terms of Lagrangians. Only a small part of interacting superconformal field theories are known to have a Lagrangian description, and sometimes even in these cases not the entire superconformal structure is seen in the Lagrangian. The now classical example is the ABJM theory \cite{ABJM} in three dimenions with Chern-Simons levels $|k|=1,2$. In this particular case only the ${\cal N}=6$ part of the entire ${\cal N}=8$ superconformal structure is seen in the Lagrangian.  

Another, more general way to define a superconformal field theory is as the infrared fixed point of some supersymmetric QFT which is usually a perturbation of an UV free fixed point by a relevant operators. In this case we know, for example, the symmetries\footnote{Except the so-called 'accidental symmetries' whose currents are not conserved along the entire RG flow but only in the infrared.} of the IR superconformal field point but not its Lagrangian. In fact there is no reason to believe that such a Lagrangian should exist in general.

This makes it clear that classification of superconformal quantum field theories is not reduced to the classification of the superconformal Lagrangians. To find general properties of superconformal theories more abstract approach not relying on the possibility of a Lagrangian description should be employed. In this note we make use of only the most fundamental characteristics of ${\cal N}\ge 6$ superconformal field theories in three dimensions to find and prove some of their properies. These characteristics are: existence of the superconformal algebra, unitarity and the existence of the stress-tensor.

In the recent years many ${\cal N}=6$ and ${\cal N}=8$ superconformal quantum field theories were found in three space-time dimenions \cite{BL,Gustavsson,ABJM, ABJ}. All these theories have an interesting property -- they contain a global $U(1)$ symmetry. In the first part of this note we explain this 'empirical' fact as stemming only from the properties of ${\cal N}=6$ superconformal algebra, unitarity and the existence and uniquess of the stress-tensor. As a result, every 'irreducible' \footnote{That is, possessing a unique stress-tensor.}${\cal N}=6$ superconformal quantum field theory contains a single conserved global current. This immediately implies that any ${\cal N}=8$ superconformal theory has no global symmetries. In the second part of the note we make use of this result to explain another 'empirical' fact -- the fact than no purely ${\cal N}=7$ superconformal field theories have been found so far. We show that every ${\cal N}=7$ superconformal field theory has actually a larger, ${\cal N}=8$, supersymmetry.

\section{${\cal N}=6$ superconformal field theories}  

We start by reviewing the structures of the stress-tensor and global conserved currents multiplets in ${\cal N}\ge 4$ three dimensional superconformal field theories.
In three dimensional ${\cal N}$ superconformal field theory the stress-tensor is a primary field with conformal dimension three which is spin-two tensor $T_{(\alpha\beta\gamma\delta)}$ with respect to the rotation group $SO(3)$\footnote{We work with the Euclidean version of the theory.} of ${\mathbb R}^3$ and a singlet of the $SO({\cal N})_R$ $R-$ symmetry and any global symmetry groups. This tensor belongs to a supermultiplet ${\cal T}$ which can be decomposed with respect to the bosonic subgroup $SO(2)\times SO(3)\times SO({\cal N})_R$ of the superconformal group as ${\cal T}=\oplus_{n,j,R}{\cal T}_{n,j,R}$ where $n$ denotes the so-called level and runs from zero to infinity. As $n$ increases by one the conformal dimension of the representations living on the same level increases by one-half. The only operators from the superconformal algebra that raise or lower the level are supercharges $Q$ and their conjugates (in the radially quantized picture) superconformal charges $S$ as they are the only operators that do not commute with the dilatation operator. Indices $j$ and $R$ label the spins and $SO({\cal N})_R$ representations. The lowest level $n=0$ contains a single representation of $SO(3)\times SO({\cal N})_R$ with spin $j_0$, with the highest weight $(r_1,...,r_{[{\cal N}]/2})$ of $SO({\cal N})_R$\footnote{We use the convention $r_1\ge r_2\ge...\ge r_{[{\cal N}/2]}$} and the conformal dimension $\epsilon_0$ subject to a certain inequality stemming from the requirements of unitarity \cite{Minwalla}.

The lowest component of the stress-tensor multiplet is an $SO(3)$ scalar and absolutely antisymmetric rank four $SO({\cal N})_R$ tensor\footnote{In the case of ${\cal N}=8$ the antisymmetric rank four tensor is decomposed into a self-dual and anti-selfdual parts. Choosing either of them is a matter of convention.} with conformal dimension $\epsilon_0=r_1=1$. On the second level of the stress-tensor multiplet are the $R$-currents, on the third level are the supercurrents and on the fourth level is the stress-tensor itself.   

Similarly, a conserved global current, if it exists, belongs to a supermultiplet. The lowest component of this supermultiplet is an $SO(3)$ scalar which is the rank-two antisymmetric tensor of $SO({\cal N})_R$ with conformal dimension $\epsilon_0=r_1=1$ \cite{GW}.

Unless ${\cal N}=6$ the stress-tensor and a global current multiplets are two distinct multiplets. When ${\cal N}=6$ a rank-two antisymmetric tensor is equivalent to a rank-four antisymmetric tensor. This means that the stress-tensor multiplet may contain a conserved global $U(1)$-current on the second level. Below we argue that this is indeed the case: every ${\cal N}=6$ superconformal field theory has a global $U(1)$ symmetry whose current lives on the second level of the stress-tensor multiplet together with the $R$-currents.

First of all we note that all known ${\cal N}=6$ superconformal field theories possess a global $U(1)$ symmetry. In the case of ABJM theories with gauge groups $SU(N)\times SU(N)$ the $U(1)$ symmetry is the barion number symmetry, while in the case of the gauge group $U(N)\times U(N)$ it is the symmetry generated by the topological current $J=\star\frac{1}{4\pi}(tr(F)-tr(\tilde F))$.

Now consider a simple case of the ABJM theory \cite{ABJM} with gauge group $U(1)_k\times U(1)_{-k}$ where the subscipts stand for the Chern-Simons levels. In addition to the two ${\cal N}=2$ vector multiplets there are matter fields -- two hypermultiplets in the bifundamental representation of the gauge group $U(1)\times U(1)$. In terms of the chiral multiplets these are $(A_1, B_1)$ and $(A_2,B_2)$. The theory contains a quartic superpotential and the Chern-Simons kinetic term for the gauge fields with Chern-Simos level $k$ for the first $U(1)$ and $-k$ for the second $U(1)$. This theory has ${\cal N}=6$ supersymmetry unless $k=1,2$\footnote{If $k=1$ or $k=2$ the supersymmetry is enhanced to ${\cal N}=8$.}. Moreover, as we mentioned, there is a $U(1)$ global symmetry. The rank-two antisymmetric tensor of $SO(6)_R$ is equivalent to a rank-four antisymmetric tensor. This is the representation ${\bf 15}$ of $SO(6)_R$. Because there is a $U(1)$ global symmetry we must have another ${\bf 15}$ with conformal dimension one in addition to that corresponding to the stress-tensor. However, we only see one copy of ${\bf 15}$ with conformal dimension one\footnote{Due to a large amount of supersymmetry (${\cal N}>2$) all fields have their UV conformal dimensions.} -- the binomials in the matter scalars $C_IC^\dagger_J$ where $C_I=(A_1,A_2, B_1^\dagger, B_2^\dagger)$ is in the spinor representation ${\bf 4}$ of $SO(6)_R$. There are no candidates for another copy of ${\bf 15}$. 

Actually, it is easy to prove that there is no other ${\bf 15}$.  Let us consider only chiral scalars -- those scalars which are annihilated by a complex supercharge $Q$ corresponding to an $SO(2)$ subgroup of $SO(6)$. The representation ${\bf 15}$ is decomposed as ${\bf 15}={\bf 6}_0+{\bf 4}_1+{\bf 4}_{-1}+{\bf 1}_0$ under $SO(4)\times SO(2)\subset SO(6)_R$. The representation ${\bf 4}_1$ consists of four chiral scalars. Now we can compute the superconformal index for these theories on $S^2\times{\mathbb R}$\cite{BBMR}, that is, in the radially quantized picture
\begin{align}
{\cal I}(x)=Tr[(-1)^Fx^{\epsilon+j_3}]
\end{align}
using the localization technique \cite{Kim}. In the above expression $F$ is the fermion number. 

In the Taylor expansion of the index around $x=0$ the coefficient in front of the first power of $x$ counts the number of chiral scalars with conformal dimension one. No other states can contribute to this coefficient because of the unitarity constraints. If the coefficient is four, there are only 4 chiral scalars and, correspondingly, only one representation ${\bf 15}$ of $SO(6)_R$. If the coefficient is eight, there are two copies of ${\bf 15}$. Of course, the coefficient does not depend on the particlular choice of an ABJM theory.  We computed the index for the $U(N)_k\times U(N)_{-k}$ theory for several low values of $k>2$ and $N$ and found
\begin{align}
{\cal I}(x)=1+4x+{\cal O}(x^2)
\end{align} 
The conclusion is that there is only one representation ${\bf 15}$ of $SO(6)_R$ which is an $SO(3)$ singlet with conformal dimension $\epsilon_0=1$. This means that both the $R$-currents and the $U(1)$ symmetry current are obtained from the same fifteen scalars by acting on them with a bilinear combination of supercharges $Q_{(\alpha}^{[i} Q_{\beta)}^{j]}$. Here latin indices are fundamental indices of $SO(6)_R$ while the greek indices are spinor indices corresponding to space-time rotations. The group theory indeed allows that because ${\bf 15}\times {\bf 15}={\bf 1}+{\bf 15}+...$. Note that the group theory argument alone is insufficient because the norm of the $SO(6)_R$ singlet could turn out to be zero and this state then would be absent in the Hilbert space of the radially quantized theory or as an operator on ${\mathbb R}^3$. Our example shows that this is not the case. The $SO(6)_R$ singlet current does exist and is conserved by virtue of its quantum numbers. 

The conclusion about the existence of a global $U(1)$ symmetry is in fact true for any ${\cal N}=6$ superconformal quantum field theory. Indeed, the norm of this $SO(6)_R$ singlet current which is obtained from $Q_{(\alpha}^{[i} Q_{\beta)}^{j]}\ket{\bf 15}$ where ${\bf 15}$ are the scalars on the zero level of the stress-tensor supermultiplet  is determined by only the superconformal ${\cal N}=6$ algebra. So, in any ${\cal N}=6$ superconformal theory the $U(1)$ current has a non-zero norm. Furhermore, if there is more than one global symmetry in the theory, then there is more than one set of associated scalars in the representation ${\bf 15}$ of $SO(6)_R$. Each such set generates the whole stress-tensor supermultiplet (to which the global current belongs). In particular, there is more than one stress-tensor. Such a theory is reducible, i.e. decomposes into a direct sum of several superconformal field theories. Thus the presence of more than one $U(1)$ global symmetry is an indicator of 'reducibility' of ${\cal N}=6$ SCFT. There are in fact examples of 'reducible' superconformal quantum field theories both in four dimensions \cite{GST} and in three dimensions \cite{BK,BK2} where 'reducibility' is not obvious in the (UV) Lagrangian description.

We come to the main conclusion of this section -- every irreducible ${\cal N}=6$ superconformal theory has a single global $U(1)$ symmetry with its current appearing on the second level of the stress-tensor supermultiplet together with the $SO(6)_R$ currents. 

\section{${\cal N}=7$ superconformal field theories}

Now that the existence of a global $U(1)$ symmetry for any ${\cal N}=6$ superconformal field theory is established there arises a natural question -- how the global $U(1)$ symmetry fits into the cases of ${\cal N}=7$ and ${\cal N}=8$ superconformal symmetries?\footnote{I thank Anton Kapustin for asking me this natural question.} For the case of ${\cal N}=8$ the answer is obvious -- just as in the case of ABJM theories  \cite{ABJM} the global $U(1)$ symmetry becomes  the commutant of $SO(6)_R$ in the full $R$-symmetry group $SO(8)_R$. Because an irreducible ${\cal N}=8$ superconformal theory is an irreducible ${\cal N}=6$ superconformal theory with a global (with respect to the ${\cal N}=6$ superconformal structure) $U(1)$ symmetry which corresponds to the commutant of $SO(6)_R$ in $SO(8)_R$, there is no room for any global symmetries. Thus any ${\cal N}=8$ SCFT has no global symmetries. In this section we explore the way in which the structure of ${\cal N}=7$ theories is affected. We find that there are no purely ${\cal N}=7$ superconformal field theory -- every ${\cal N}=7$ superconformal theory is in fact ${\cal N}=8$ supersymmetric.

Because an ${\cal N}=7$ superconformal field theory is a particular case of ${\cal N}=6$ superconformal field theories there is a $U(1)$ symmetry which is global as long as the ${\cal N}=6$ subgroup of the ${\cal N}=7$ supergroup is considered. There are two options for the $U(1)$ group to fit into the $SO(7)_R$ $R$-symmetry group. First is that the $U(1)$ does not commute with the $SO(7)_R$. This immediately implies that the full $R$-symmetry group is $SO(8)$ and so the theory is ${\cal N}=8$ supersymmetric.

The second option is that the $U(1)$ commutes with the $SO(7)_R$. Let us check if this option is self-consistent. The lowest components (operators on ${\mathbb R}^3$) of the stress-tensor supermultiplet are scalars with the conformal dimension one and form a fourth-rank antisymmetric tensor with respect to $SO(7)$. This tensor is equivalent to a third-rank antisymmetric tensor. It is easy to see how this representation decomposes with respect to $SO(6)_R\subset SO(7)_R$: ${\bf 35}={\bf 15}+{\bf 10}+{\bf \bar{10}}$. This is exactly how the lowest component of the stress-tensor supermultiplet of an ${\cal N}=8$ theory decomposes under $SO(6)_R\subset  SO(8)_R$ \cite{BK}.

Here we remind the reader that the lowest component of the stress-tensor supermultiplet of an ${\cal N}=8$ superconformal theory is a rank-four selfdual antisymmetric tensor\footnote{Or an antiselfdual tensor -- this depends on the convention.} ${\bf 35}$. Under the reduction $SO(8)_R\to SO(7)_R$ it becomes a single irreducuble representation ${\bf 35}$ of $SO(7)$. 

If the global symmetry $U(1)$ commutes with $SO(7)_R$ then all three irreps ${\bf 15}+{\bf 10}+{\bf \bar{10}}$ of $SO(6)$ have zero $U(1)$ charges. In the case of ${\cal N}=8$ it was ${\bf 15}+{\bf 10}_1+{\bf {\bar 10}}_{-1}$ instead. Forgetting for a moment about the $U(1)$ charges consider the action of the supercharges bilinears $Q_{(\alpha}^{[i} Q_{\beta)}^{j]}$ on the scalars in the representation ${\bf 10}+{\bf \bar{10}}$ where $i,j$ are fundamental indices of $SO(6)_R$ and greek indices are rotation spinor indices. In \cite{BK} it was shown that the result of this operation is the set of conserved currents in the representation ${\bf 6}$ of $SO(6)_R$ needed to enhance $SO(6)_R\times U(1)$ to $SO(8)_R$. Namely, the currents are in the representation ${\bf 15}+{\bf 6}+{\bf 6}$ of $SO(6)_R$, where the representation ${\bf 15}$ comes form $Q_{(\alpha}^{[i} Q_{\beta)}^{j]}\ket{\bf 15}$. For ${\cal N}=7$ the conserved currents in the representation ${\bf 6}$ of $SO(6)_R\subset SO(7)_R$ are the currents which enlarge $SO(6)_R$ to $SO(7)_R$: ${\bf 21}={\bf 15}+{\bf 6}$. However there are twice as many of them as needed for $SO(7)_R$. Thus there are conserved currents (with nonzero norm) in addition to those in $SO(7)_R$ which do not commute with the $SO(7)_R$. This means that the supersymmetry is enhanced to ${\cal N}=8$ which contradicts the assumption that the theory has only ${\cal N}=7$ superconformal symmetry.

This proves the claim that every ${\cal N}=7$ superconformal theory is in fact ${\cal N}=8$ supersymmetric. 

\section{Acknowledgments} I would like to thank Anton Kapustin for reading the note and making useful suggestions.

\end{document}